%% file: main.tex
\documentclass[10pt,conference]{IEEEtran}
\IEEEoverridecommandlockouts
\usepackage{cite}
\usepackage{amsmath,amssymb,amsfonts}
\usepackage{algorithmic}
\usepackage{graphicx}
\usepackage{textcomp}
\usepackage{xcolor}
\usepackage{booktabs}
\usepackage{hyperref}
\usepackage{framed}
\def\BibTeX{{\rm B\kern-.05em{\sc i\kern-.025em b}\kern-.08em
    T\kern-.1667em\lower.7ex\hbox{E}\kern-.125emX}}

\definecolor{pblue}{rgb}{0.13,0.13,1}
\definecolor{pgreen}{rgb}{0,0.5,0}
\definecolor{pred}{rgb}{0.9,0,0}
\definecolor{pgrey}{rgb}{0.46,0.45,0.48}

\usepackage{listings}
\begin{document}

%\title{Combining Code Embedding with Static Code Analysis for Function Completion}
\title{Combining Code Embedding with Static Analysis for Function-Call Completion}

\author{
\IEEEauthorblockN{Martin Weyssow}
\IEEEauthorblockA{DIRO, Université de Montr\'eal\\
Montreal, Canada\\
martin.weyssow@umontreal.ca}
\and
\IEEEauthorblockN{Houari Sahraoui}
\IEEEauthorblockA{DIRO, Université de Montr\'eal\\
Montreal, Canada\\
sahraouh@iro.umontreal.ca}
\and
\IEEEauthorblockN{Beno\^it Fr\'enay}
\IEEEauthorblockA{University of Namur \\
Namur, Belgium \\
benoit.frenay@unamur.be}
\and
\IEEEauthorblockN{Beno\^it Vanderose}
\IEEEauthorblockA{University of Namur \\
Namur, Belgium \\
benoit.vanderose@unamur.be}
}

\maketitle

\begin{abstract}
Code completion is an important feature of integrated development environments (IDEs). It allows developers to produce code faster, especially novice ones who are not fully familiar with APIs and others’ code. Previous works on code completion have mainly exploited static type systems of programming languages or code history of the project under development or of other projects using common APIs. In this work, we present an approach for improving current function-calls completion tools by learning from independent code repositories, using well-known natural language processing models that can learn vector representation of source code (\textit{code embeddings}). Our models are not trained on historical data of specific projects. Instead, our approach allows to learn high-level concepts and their relationships present among thousands of projects. As a consequence, the resulting system is able to provide general suggestions that are not specific to particular projects or APIs. Additionally, by taking into account the context of the call to complete, our approach suggests function calls relevant to that context. We evaluated our approach on a set of open-source projects unseen during the training. The results show that the use of the trained model along with a code suggestion plug-in based on static type analysis improves significantly the correctness of the completion suggestions.
\end{abstract}


\section{Introduction}
    \input{content/1_introduction}

\section{Background}
    \input{content/2_background}

\section{Our Approach}
    \input{content/3_our_approach}

\section{Evaluation Setup}
    \input{content/4_evaluation_setup}

\section{Evaluation Results}
    \input{content/5_evaluation_results}

\section{Related Work}
    \input{content/6_related_work}

\section{Conclusion}
    \input{content/7_conclusion}

\bibliographystyle{IEEEtran}
\bibliography{references}

\end{document}

%% file: content/1_introduction.tex
\label{sec:introduction}

Nowadays, developers rely on features provided by modern Integrated Development Environments (IDEs) to ease their cognitive load and increase their productivity. One purpose of these features is to avoid asking developers to provide information that can be inferred from the available data sources and the current development context \cite{murphy_context}. Among these features, code completion is one of the most widely used by, among others, Java developers in Eclipse \cite{Murphy_eclipse}. Code completion helps developers to write code faster by providing a list of suggestions of possible elements, such as function calls, pertinent to a given context.

There have been a lot of research contributions that attempt to improve code completion systems. Early learning-based approaches focused on completion, specifically for APIs by leverage historical or context data about the system under development \cite{bruch2009,proksch2015}. From another perspective, work has been done to exploit natural language modeling for, among other tasks, code completion, based on the idea of code naturalness \cite{naturalness, localness, ngram_cache}.
More recently, other approaches have targeted AST representations of the code to perform the APIs calls completion \cite{gralan, bhoopchand2016, li2017, Svyatkovskiy_2019, svyatkovskiy2020fast}. In general, the above-mentioned works exploit historical data from the projects used during the evaluation of the system and/or evaluate their systems on specific APIs completion. In the first case, the approaches are not applicable to new projects or projects with short histories, whereas, in the second  case, the objective is to predict the calls to APIs' methods. Although the obtained results are convincing, these approaches have shown to be efficient only for popular libraries.

In this paper, we propose an approach for improving function calls completion by learning models from independent code repositories. Our goal is to allow call completion not only with API functions, but also those of the project under development. More specifically, we consider each method as a natural text paragraph containing a sequence of function calls. Then, using a well-known word embedding model, we learn vector representation of variable-length sequences of these paragraphs. Our approach is based on the assumption that there exist recurring  patterns of function-call sequences among the code repositories and that these patterns capture semantics about higher-level concepts. Our approach is intended to abstract these high-level concepts and use them to improve function-call completion by comparing the call site context with the huge amount of contexts learned from the repositories. We use the learned models for function-call completion by combining them with a static analysis performed on the project under development. A type-based static analysis allows us to retrieve the list of possible function calls given a completion site. The goal is to rank this list using the embedding model with the most likely calls at the top.

To evaluate the proposed approach, we used a corpus of more than 14,000 Java projects from which we extracted more than 10 millions function sequences to train our models. To test our completion strategies, we selected 10 projects, not considered for the training, and having more than 160.000 call sites to complete. The results of our evaluation show, on the one hand, that the ranking of a list of possible candidates retrieved by static analysis improves the completion precision of the static analysis tool, for 9 of the 10 projects, by up to 135\% reaching 85\% Recall@10. On the other hand, given the promising results obtained, we compared our approach with state-of-the-art language models used in code completion. We considered two variants of language models: (1) a $n$-gram language model and (2) a $n$-gram model augmented with a cache component that has shown to outperform LSTM-based deep learning models in source code modeling \cite{ngram_cache}. In the former, we show that the $n$-gram model is not able to improve the ranking of the static analysis tool. While, in the latter, even though the $n$-gram model stores cache information about the system under development, the results show that our approach is much more efficient in term of Recall@10 and MRR. Finally, we found that it takes between 700 ms and 800 ms, on average, to produce completion suggestions for a call site. This makes our approach usable in a real programming setting.

The rest of the paper is structured as follows. In Section \textbf{\ref{sec:background}}, we introduce the word embedding and language models that we use in this work. Section \textbf{\ref{sec:our_approach}} presents the general approach for building a code completion system based on embedding models and explains its integration within an existing typing-based tool. We present the evaluation setup in Section \textbf{\ref{sec:evaluation_setup}} and report on the results in Section \textbf{\ref{sec:evaluation_results}}. Later, we discuss the related work in Section \textbf{\ref{sec:related_work}}. Finally, we draw conclusion and list future work directions in Section \textbf{\ref{sec:conclusion}}.

%% file: content/2_background.tex
\label{sec:background}

In this section, we review $n$-grams language models. Then, we describe word embedding models to learn vector representations of variable-length texts.
In Section \textbf{\ref{sec:exp_ngrams}}, we use $n$-grams to quantify the predictability of function sequences in source codes. In Sections \textbf{\ref{sec:exp_pvmodel}} and \textbf{\ref{sec:exp_comparison}}, we compare $n$-gram-based and word embedding approaches for function-call completion. 

\subsection{$n$-gram Language Models}
    Language models (\textit{LMs}) assign probabilities to sequences of words.  The main purpose of such model is to capture regularities in a large training corpus as leverage to some downstream task(s) (\textit{e.g., speech recognition, spelling correction, text generation...}). 
    
    Considering a word sequence $w_1, w_2, ..., w_n$, a LM assigns a probability $P(w_1, w_2, ..., w_n)$ to the sequence. Such a probability is hard to compute because usually long sequences of words are not observed in a training corpus. Therefore, we use $n$-gram language models to approximate $P(w_1, w_2, ..., w_n)$.
    
    $n$-gram language models assign a probability to a word $w$ given an history of size $n - 1$. $n$-gram LMs assume that the occurrence of a word depends only on the previous words. In other words, a $n$-gram model is a \textbf{Markovian approximation} of order $n - 1$:
    $$
        P(\mathbf{w}_1^n = w_1, w_2, ..., w_n) \approx \prod_{k=1}^{n} P(w_k | w_{k-n+1}^{k-1}).
    $$
    
    The simplest approach to estimate these word probabilities is a maximum likelihood estimation (\textbf{MLE}) over the raw counts of words in the corpus. In practice, MLE is not used to avoid the model to assign zero probability to unseen sequences of words. Instead, smoothing techniques are used and assign part of the total probability mass to unseen $n$-grams. For instance, \textbf{Kneser-Ney} is one of the most known smoothing techniques and is very efficient \cite{kneser-ney-1995, chen-goodman-1996}.
    
    \subsubsection{Language model evaluation}
    A good language model predicts with a low-level of uncertainty the content of an unseen piece of text. The level of uncertainty of a language model can be measured by the \textbf{cross-entropy}. Given a $n$-gram language model $L$ and a word sequence $\mathbf{w}_1^n = w_1, w_2, ..., w_n$, the cross-entropy is computed as:
    $$
        H_L\left(\mathbf{w}_1^n\right) = - \frac{1}{n} \sum_{i=1}^{n} \log P(w_i | w_{i-n+1}^{i-1}).
    $$
    For the case of a $n$-gram model, the cross-entropy is the average number of bits required to predict the $n^{th}$ word given the $n-1$ previous words. Consequently, a model that has low entropy on a given piece of text has a low-level of uncertainty and predicts with confidence the content of the text.

\subsection{Distributed Representations of Words}
    \textbf{Word embedding} is a technique commonly used in natural language processing (\textit{NLP}) to learn a mapping of words into an high-dimensional vector space. The notion of word embedding is highly related to \textbf{distributional semantics}. That is quantifying some semantic similarities between words or concepts that appear frequently in the same context in a large corpus of textual data. Two words that have close vector representation are meant to be semantically similar. For example, it is likely that $senate$ and $politic$ would be close. 
    
    One of the most-known framework for learning distributed representation of words is \textbf{Word2vec} \cite{word2vec_1, word2vec_2}. Nevertheless, there is no inherent scheme to the model to learn embedding of sequences of words. Such an approach would, for example, allow us to compute the similarity between two text documents (\textit{e.g., of variable-length}). Paragraph vector model aims to tackle this problematic by learning vector representation of variable-length texts.

\subsection{Paragraph Vector Embedding Model}

    The paragraph vector (\textit{PV}) model, \textit{i.e.,} \textbf{Doc2vec}, is an extension of Word2vec proposed by Le and Mikolov \cite{doc2vec}. PV models learn vector representations (\textit{paragraph vectors}) of sequences of textual data of variable size (\textit{document, phrases, news article...}). In this model, each input sequence has a unique corresponding paragraph vector that is learned along with the word vectors. Paragraph vectors are not just concatenation and average of word vectors contained within the paragraph. Instead, paragraph vectors are asked to contribute to a predictive task as for words in Word2vec.

    % There exist two architectures for the learning process : distributed memory (\textbf{PV-DM}) and distributed bag-of-words (\textbf{PV-DBOW}). 
    % \begin{itemize}
    %    \item In PV-DM, the model randomly sample contexts within the paragraph. The contexts are determined by a window size. Then, an average or concatenation of the paragrah vector and the word embeddings is used to predict the last word of the sampled context. 
    %    Figure \textbf{\ref{fig:pv-dm}} illustrates this architecture.

    %    \begin{figure}[!ht]
    %        \centering
    %        \includegraphics[width=0.4\textwidth]{images/pv-dm_architecture.png}
    %        \caption{PV-DM. The context is randomly sampled using a window size of 4. Words and paragraph embeddings are averaged or concatenated to predict the target word.}
    %        \label{fig:pv-dm}
    %     \end{figure}
        
    %    \item In PV-DBOW, the model sample contexts similarly to PV-DM. However, context words are ignored in the input. The paragraph vector is asked to predict random words from the sampled context. 
    %    This is illustrated in Figure \textbf{\ref{fig:pv-dbow}}.
    %\end{itemize}
    %\begin{figure}[!ht]
    %    \centering
    %    \includegraphics[width=0.2\textwidth]{images/pv-dbow_architecture.png}
    %    \caption{PV-DBOW. The context is ignored in the input. The paragraph vector is used to predict the context words. }
    %    \label{fig:pv-dbow}
    %\end{figure}
    
    The advantage of the PV model over Word2vec is that the model is able to learn representation of variable-length texts. As result of the learning phase, the paragraph vectors can capture semantic properties about whole sequences. This model has shown to be useful in topic modeling and several NLP tasks \cite{pv_topic-detection1, pv_topic-detection2, doc2vec_empirical}.

%% file: content/3_our_approach.tex
\label{sec:our_approach}

To illustrate the rationale behind our approach, let us consider the situation in which, Ulwazi, a Java developer, is writing the method in Listing \textbf{\ref{lst:motivation}}.

\begin{minipage}{\linewidth}
\begin{lstlisting}[caption={Motivating example}, label={lst:motivation},captionpos=b]
// ...
public long size() throws IOException {
  if (!file.isFile()) {
    throw new FileNotFoundException(
      file.toString());
  }
  return file.? // prediction (ctrl+space)
}
\end{lstlisting}
\end{minipage}

Consider also that Ulwazi is coding in an IDE that incorporates, among other features, a code completion plug-in such as Eclipse content assist\footnote{https://www.eclipse.org/documentation/} that suggests function calls. 
In line 7, after she types " . " the plug-in is invoked, and the latter will provide a suggestion list of possible items including function calls that could follow "\textit{file.}". The plug-in exploits static environment information about the currently opened code artefact (\textit{e.g., imports, language typing...}). The produced suggestion list is generally exhaustive, often long, and ordered alphabetically. Thus, it is more likely that the correct suggestion will not appear at the top of the list, and developers like Ulwazi will waste a valuable time browsing through the list. 

Therefore, our objective is to alleviate the burden of developers by providing completion suggestion lists that are: (1) of limited size, and (2) ordered by pertinence so that the correct suggestion is likely to appear in the top positions. 

Our approach is based on the hypothesis that there exist recurring function-call patterns in large corpus of source code. Those patterns embody some semantics about high-level concepts, which may appear in different programs with slight linguistic variations. 
For instance, coming back to the example of the method "\textit{size}" that computes the size of a file, the first step consists in checking whether the input is a file, by calling, for example, a function "\textit{.isFile()}". If it is not, one may want to raise an exception with a representation of the file by calling "\textit{.toString()}". The final step is to call a function "\textit{.length()}" that outputs the size of the file. Our approach makes the assumption that such sequences of function calls are totally or partially recurrent among a lot of projects and that they capture most of the semantics of some higher-level concepts (\textit{in our case, "get the size of a file}"). By comparing the previous function call sequence (\textit{including the method name}) "\textit{size, isFile, toString}" with
function call sequences abstracted by a word embedding model, it would be possible to determine that "\textit{length}" is the most probable function call that comes after "\textit{file.}".

\begin{figure*}[!t]
    \centering
    \includegraphics[width=.95\textwidth]{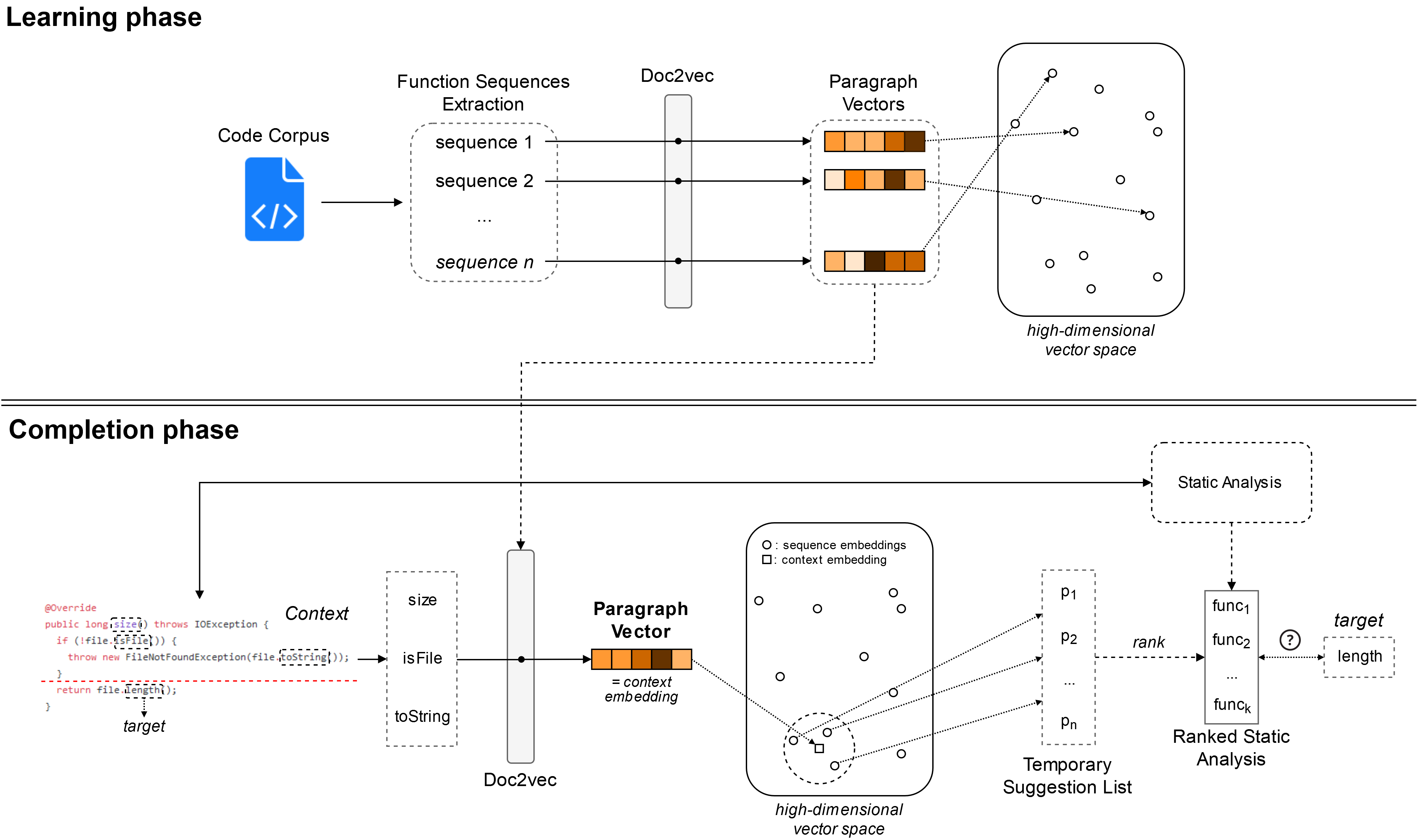}
    \caption{Approach - General framework}
    \label{fig:approach}
\end{figure*}

In this work, we propose an approach to learn those high-level concepts and their relationships by training an embeddings model (\textit{e.g., paragraph vector model}) on a big corpus of code. Once the model trained, we can take advantage of it for the function completion task by ordering the list of possible function calls retrieved with a type-based static analysis tool. We describe the learning process in Section \textbf{\ref{sec:approach_learning}} and the function completion in Sections \textbf{\ref{sec:approach_completion}}. Figure \textbf{\ref{fig:approach}} illustrates both processes.

\subsection{Learning Concepts from Code}
\label{sec:approach_learning}

As discussed in the previous section, sequences of function calls embody a great part of the semantics of code. Therefore, it might be a good way to use these sequences as textual representations of the source code. Conversely, syntactical tokens such as \textit{if, else, for} or a \textit{parenthesis} carry less domain-specific information and considering them could lead to introducing a lot of noise in the learning, especially in the context of function call completion. It is also important to define how do we cut the code in order to produce sequences of functions and learn the paragraph vectors. We propose to limit the scope of a sequence to a method declaration and its body as for the \textit{size} method. A method can be seen as a paragraph that is designed to deal with a particular concern, as we would do in a text. And, it is more likely that functions sequences within a small scope are more recurring than in a broader scope, \textit{e.g.,} a whole class.  Furthermore, limiting the sequences to a relatively small scope allows the model to learn specific and precise concepts. 

Our approach is also not limited to using sequences of functions. One alternative would be to consider subtokens of function names instead of the full functions names. Given a function name, the subtokens are words contained within it. For example, if we have a function that is called "\textit{convertDateToString}", we tokenize the camel case and the resulting subtokens are "\textit{convert,\: date,\: to,\: string}". This approach has been used in previous works on source code modeling \cite{Allamanis2016_convolutional, Allamanis_suggestion, karampatsis2020big, svyatkovskiy2020fast}. It has shown to be useful to summarize code snippets and for suggesting out-of-vocabulary code tokens (\textit{names that does not appear during the training phase of a model}).

The learning process is described in the upper part of Figure \textbf{\ref{fig:approach}}. The first step extracts function sequences from a corpus of code. Then, the sequences are used as input of a paragraph vector model (\textit{Doc2vec}). Finally, the model learns high-dimensional vector representations of the function sequences (\textit{paragraph vectors}). 

\subsection{Function Completion with Static Analysis}
\label{sec:approach_completion}
In most recent works on code completion, the authors proposed different approaches where a suggestion list of most-likely candidates is built for a specific context using a standalone model \cite{Liu2017NeuralCC, karampatsis2020big, ngram_cache, kim2020code}. The main drawback of such approaches is that they do not guarantee that all the suggestions are feasible in the current project. In this work, we propose an approach where, instead of building a list of suggestions with a standalone model, we rank the list of all possible function calls retrieved using a type-based static analysis tool with a learned model. The advantage of this approach is that the static analysis provides functions depending on the packages/libraries imported in the code file in which the method is implemented. Also, it filters the calls that are not type compliant \textit{w.r.t} the completion site. Therefore, this approach guarantees that the suggestion list will only contain tangible function calls. 

The completion process is described in the lower part of Figure \textbf{\ref{fig:approach}}. Again, we consider the scenario of Section \textbf{\ref{sec:our_approach}} in which Ulwazi is implementing a class method. The completion process is designed in five steps :

\begin{enumerate}
    \item \textbf{Extraction of the context}. The context is made of the  name of the 
    method under development followed by the sequence of calls, in the method body, preceding the call site that triggers the completion. If the call to complete is the first of the method, then the context is only made of the name of the method.
    
    \item \textbf{Static analysis}. A static analysis is performed to retrieve all possible function calls given the context of the completion site.
    
    \item \textbf{Inferring a paragraph vector}. Using the previously trained model, we infer a vector representation of the context (\textit{context embedding}). 
    
    \item \textbf{Building a temporary suggestion list}.
    We use the embedding of the context to retrieve the closest paragraph vectors in the model. This can be done by finding the paragraph vectors that have the greatest cosine similarity to the context vector. Since paragraph vectors correspond to sequences of functions, this is like retrieving the most similar sequences of functions to the context. Then, we build a temporary suggestion list using the retrieved sequences. We add functions of the retrieved sequences to the suggestion list and stop when the list has reached its maximum allowed size or a threshold of the similarity score.
    
    \item \textbf{Ranking the static analysis}. Using the temporary suggestion list, we rank the possible function calls provided by the static analysis. We iterate over the functions in the temporary list. For each function, if it appears in the static analysis then we add it to the final suggestion list. We stop the process when the list has reached certain size or when the PV model is not able to find more similar sequences to the context.
    
\end{enumerate}

As we will see in the evaluation, the steps 3, 4 and 5 can be generalized to any model that is able to provide a list of candidates for a completion site. This allows us to compare our approach using a paragraph vector model with state-of-the-art $n$-gram language models.

%% file: content/4_evaluation_setup.tex
\label{sec:evaluation_setup}

Previous works have shown that source code is (\textit{locally}) repetitive and predictable using statistical language models \cite{naturalness, localness}. Recent works have found that variable and function identifiers are the main responsible for the high-level of entropy of code and that syntax tokens artificially increase the source code predictability \cite{naturalness_revisited}. Thus, one of the key challenges of learning high-level concepts from codes using sequences of functions lies in the high-level of unpredictability of those sequences. This leads us to address the following research questions:

\begin{itemize}
    \item \textbf{RQ1 [Replication]: \textit{How repetitive and predictable are function sequences in source code?}} \\
    We reproduce previous works on naturalness of software \cite{naturalness, naturalness_revisited}. We check whether our datasets satisfy the naturalness hypothesis introduced by \textit{Hindle et al.} \cite{naturalness}. Then, we ensure that our datasets have a level of cross-entropy in the same order of magnitude than in \textit{Rahman et al}'s experiments \cite{naturalness_revisited}. To estimate $n$-gram language models we use SLP, a toolkit that provides fast estimation and manipulation of $n$-gram models\footnote{https://github.com/SLP-Team/SLP-Core}. As a first step, we estimate $n$-gram language models using our training sets for $n \in \left[2, 10\right]$. Then, we compute the cross-entropy on our test set. \\

    % \item \textbf{RQ2: \textit{Are paragraph vector embedding models capable of capturing concepts from the code?}} \\
    % We evaluate how well the paragraph vector model captures concepts by performing relatedness tests on some functions in the vocabulary of the trained model. The idea is to check whether close functions in the model have consistent embeddings with respect to the semantics of the functions and their usage in the code.
    
    \item \textbf{RQ2: \textit{Can we use the paragraph vector model in order to rank accurately the function calls retrieved by static analysis using Eclipse JDT Core\footnote{https://www.eclipse.org/jdt/core/} ?}} \\
    The ranking is done using the process defined in Section \textbf{\ref{sec:approach_completion}}. We used Eclipse JDT Core as a static analysis tool to retrieve the possible function calls for each completion site that appears in the 10 bold projects in Table \textbf{\ref{tab:test_set}}.
    We evaluate our approach using metrics defined in Section \textbf{\ref{sec:metrics}}.
    
    \item \textbf{RQ3 : \textit{How does the paragraph vector model performs compared to state-of-the-art $n$-grams models on function completion?}} \\
    We compare the results obtained in RQ2 with state-of-the-art $n$-gram models used by \textit{Hellendoorn and Devanbu} \cite{ngram_cache}. The authors have shown that their implementations of $n$-gram models perform well for source code modeling and code prediction. Furthermore, we also compare our approach with $n$-gram model augmented with a cache component that allows to learn local information about the project under development at test time.

\end{itemize}

\subsection{Data Source}
\label{sec:data}
    We use the GitHub Java Corpus \cite{mining_repo} consisting of more than 14,000 open-source java projects collected from Github. The corpus' statistics are presented in Table \textbf{\ref{tab:github_corpus}}. 

    \input{tables/corpus}
    
    Before forming the training set, we removed 20 projects from the original corpus to build a test set. We select these projects based on their high popularity in Github and to cover a broad range of application domains. We also considered the diversity in size. Table \textbf{\ref{tab:test_set}} shows statistics for each test project, \textit{i.e.,} the number of methods declared in each test project, the total number of call sites in these methods, and the percentage of function vocabulary that appear in the training dataset. We use the whole 20 projects to answer RQ1 and limit ourselves to the 10 projects in bold to answer the remaining questions. These 10 projects allow us to test the completion for more than 160.000 call sites. 
    
    \input{tables/test_projects_statistics}
    
    For the training of the $n$-gram models and the paragraph vector model, we extract more than 10 millions function sequences from the filtered corpus. In Table \textbf{\ref{tab:train_set1}}, we specify the number of tokens and types (\textit{i.e., unique tokens}) with and without a minimum count parameter. This parameter is used with both types of models to ignore functions that occur less than a specified threshold. The ignored functions are replaced by a common token \textit{$<$unk$>$}. We can observe that when using this minimum count parameter, the number of types decreases drastically (\textit{around 7\% of types are kept}), but the total number of tokens does not decrease that much. This means that there is a significant amount of types that are not frequent among all projects and considering them in the learning phase could lead to learning a lot of noise. 
    
    \input{tables/train_set1}
    
    In addition to this first training set, we also consider a variant of the data which consists of subtokens of the function names. We use this second training set for RQ1 to investigate the impact of representing functions as word subunits for source code modeling. 
    Tokenizing the function names considerably reduces the size of the vocabulary. Furthermore, it is more likely that each subtoken of a function such as "\textit{convertDateToString}" (\textit{i.e. "convert, date, to, string"}) appears frequently in a corpus than the whole function name.
    Table~\textbf{\ref{tab:train_set2}} shows statistics of this second training set. We can observe that the number of types is significantly lower than in the first training set and that the minimum count parameter has almost no impact on the total number of tokens.

    \input{tables/train_set2}   

\subsection{Evaluating the Paragraph Vector Model}
\label{sec:eval_pv}

For this experiment (RQ2), we first retrieved the list of possible function calls for each call site of each test project using Eclipse JDT Core. Then, we evaluated each call site by extracting the context of the call $f_c$ to be predicted. The context is made of the previous calls preceded by the method name $\left(m, f_1, f_2, ..., f_{c-1}\right)$. Next, we followed the process described in Section \textbf{\ref{sec:approach_completion}} to rank Eclipse's static analysis. For the sake of evaluation, we fixed the maximum size of the suggestion lists to 10.

To tune the hyper-parameters of the PV model, we evaluated several configurations of the model on our test set with commonly used values of hyper-parameters. We found out that the following configuration works the best for our task: PV-DBOW with dimension of the embeddings of 300, a window size of 15, a threshold of 20 for minimum word counts and a hierarchical softmax as training algorithm.

\subsection{Evaluating the $n$-gram Language Model}
\label{sec:eval_ng}
For this research question (\textit{RQ3}), we adapted the steps 3, 4 and 5 of our approach (\textit{see Section \textbf{\ref{sec:approach_completion}}}). The temporary suggestion list is built using a $n$-gram model. Given a context, the model outputs the most-likely function calls that should follow that context. Then, we rank Eclipse's static analysis using the same call sites and the same process than for the PV model in RQ2. For a fair evaluation, the maximum size of the suggestion lists is also set to 10. 

We trained $n$-gram models used in \textit{Hellendoorn and Devanbu}'s work \cite{ngram_cache}. We used Jelinek-Mercer smoothing that yields the best performances in their paper. Then, we tuned the $n$-gram model order and the vocabulary cut-off value. We found that a model order of 5 and a vocabulary cut-off of 20 gives the best performances at test time. We also considered a variant of the $n$-gram model augmented with a cache component and used the same hyper-parameters configuration than for the $n$-gram model.

\subsection{Effectiveness Metrics}
\label{sec:metrics}

The evaluation aims to determine whether a learned model is able to efficiently provide good function call suggestion lists. To evaluate our systems, we consider that a set of suggestions is relevant if it reflects the user's need. That is, the suggestion list contains the correct function call that follows a given context.

To measure the relevance, we calculate two widely-used metrics, recall at $k$ (R@$k$) and the mean reciprocal rank (MRR). As there is a unique valid suggestion for each call site, R@$k$ for a test project is the number of times the expected function call appears in top-$k$ of suggestion lists divided by the number of tested call sites.

The second metric we report is MRR. The reciprocal rank is given by the inverse of the rank of the first relevant suggestion in the result of a test sample. Mean reciprocal rank for a test set $T$ is
$$
    MRR = \frac{1}{|T|} \sum_{i=1}^{|T|} \frac{1}{rank_i}
$$
where $rank_i$ is the rank of the first relevant suggestion in the $i$-th test sample. For example, if on average, the relevant function call appears at rank 2, the MRR is 0.5. 

\subsection{Replicability Package}
\label{sec:replicability}
To facilitate the replication of our experiments, we share a Github repository that includes the artifacts to train and evaluate both $n$-gram and paragraph vector models. The repository provides extensive description on how to replicate each RQ. It also contains links to download the training and test sets\footnote{\href{https://github.com/mweyssow/cse-saner}{\textbf{link may reveal authors' identities}}}.

%% file: tables/corpus.tex
\begin{table}[!t]
\caption{GitHub Java Corpus statistics \cite{mining_repo}.}
    \centering
    \begin{tabular}{ccc} \toprule
        \# Projects & LOC & Tokens \\ \midrule
        14.785 & 352.312.696 & 1.501.614.836
    \end{tabular}
    \label{tab:github_corpus}
\end{table}

%% file: tables/test_projects_statistics.tex
\begin{table}
\caption{ Test projects used in the experiments ordered by decreasing number of function calls. Coverage is the percentage of functions that appear in the training set.}
\begin{center}
    \small

    \renewcommand{\arraystretch}{1.2}
    \setlength{\arrayrulewidth}{.5pt}
    
    \begin{tabular}{lp{0.07\textwidth}p{0.08\textwidth}p{0.06\textwidth}}
        \toprule
        Name & \# Method Decl & \# Function Calls & Coverage \\
        \midrule
        aws-sdk-java & 245.430 & 1.799.530 & 76\%  \\
        hadoop-common & 46.449 & 347.093 & 88\% \\
        spring-framework & 44.433 & 332.121 & 88\% \\
        hibernate-orm & 30.867 & 278.124 & 86\% \\
        neo4j & 33.939 & 230.914 & 80\% \\
        jclouds & 24.746 & 196.070 & 85\% \\
        cassandra & 23.398 & 188.773 & 83\% \\ 
        druid & 15.674 & 123.341 & 87\% \\
        gradle & 26.913 & 120.123 & 84\% \\
        spring-security & 13.750 & 96.950 & 84\% \\
        
        \midrule
        
        \textbf{netty} & 14.326 & 72.754 & 82\% \\
        \textbf{mongo-java-driver} & 7573 & 35.836 & 84\% \\
        \textbf{twitter4j} & 2323 & 13.365 & 99\% \\
        \textbf{clojure} & 1966 & 13.020 & 94\% \\
        \textbf{antlr4} & 2222 & 11.053 & 84\% \\
        \textbf{junit} & 2522 & 8144 & 94\% \\
        \textbf{hystrix} & 1090 & 5790 & 78\% \\
        \textbf{facebook-android-sdk} & 1453 & 5689 & 80\% \\
        \textbf{android-async-http} & 198 & 675 & 90\% \\
        \textbf{game-of-life} & 37 & 128 & 64\% \\
        \bottomrule
    \end{tabular}

\label{tab:test_set}
\end{center}
\end{table}

%% file: tables/train_set1.tex
\begin{table}[!t]
\caption{Training set 1 (\textit{full function names}). Statistics with and without minimum count parameter. Tokens corresponds to the number of method declarations and function calls in the dataset. Types is the number of unique tokens.}
    \centering
    \begin{tabular}{cccc} \toprule
        & \# Function sequences & Tokens & Types \\ \midrule
        \textit{no min count} & 10.702.667 & 86.219.928 & 3.141.457 \\
        \midrule
        \textit{min count (20)} & 10.702.667 & 74.820.025 & 222.730
    \end{tabular}
    
    \label{tab:train_set1}
\end{table}

%% file: tables/train_set2.tex
\begin{table}[!t]
\caption{Training set 2 (\textit{functions subtokens}).}
    \centering
    \begin{tabular}{cccc} \toprule
        & \# Function sequences & Tokens & Types \\ \midrule
        \textit{no min count} & 10.702.667 & 183.334.996 & 165.110 \\
        \midrule
        \textit{min count (5)} & 10.702.667 & 183.160.006 & 71.460
    \end{tabular}
    \label{tab:train_set2}
\end{table}

%% file: content/5_evaluation_results.tex
\label{sec:evaluation_results}
In this section, we present the results of our experiments and answer the research questions. For the sake of clarity, we present the global results for questions $RQ2-3$ in Table \textbf{\ref{tab:full_results}}.

\subsection{Naturalness of Function Calls (\textit{RQ1})}
    \input{content/5_1_ngram}

\subsection{Function-Call Completion with PV Model (\textit{RQ2})}
    \input{content/5_2_pvmodel}

\subsection{Comparison of PV Model and $n$-gram Language Models for Function-Call Completion (\textit{RQ3})}
    \input{content/5_3_comparison}

\subsection{Threats to Validity}
    \input{content/5_4_threats}

%% file: content/5_1_ngram.tex
\label{sec:exp_ngrams}

\begin{figure}[!t]
    \centering
    \includegraphics[width=\linewidth]{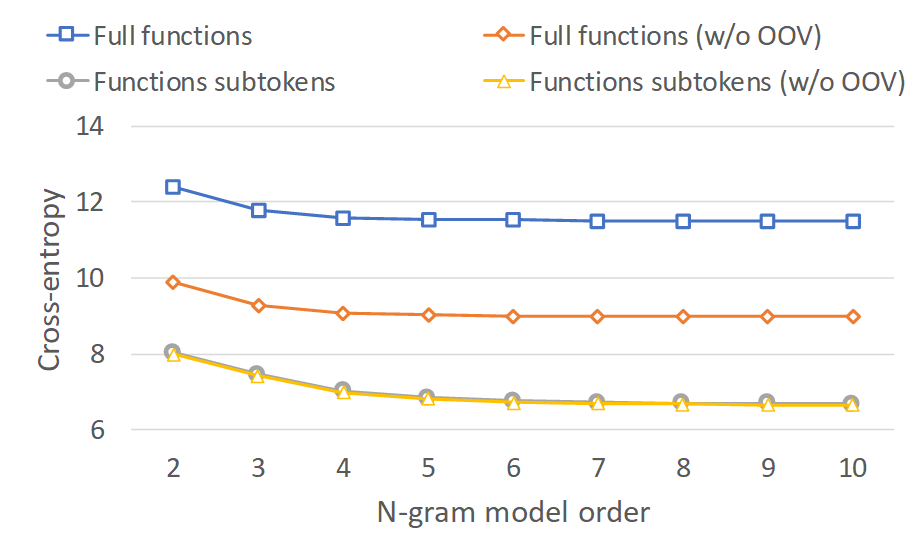}
    \caption{Comparison of the average cross-entropy on the 20 test projects for full function names and names' subtokens with respect to the order of the $n$-gram model.}
    \label{fig:ngram_global}
\end{figure}

Figure \textbf{\ref{fig:ngram_global}} shows the average cross-entropy on the 20 test projects including and excluding out-of-vocabulary (\textit{OOV}) functions. The $x$ axis represents the $n$-gram model order ($n \in [2, 10]$), which defines the size of the context considered by the model to produce a prediction. A high cross-entropy means that the next token is difficult to predict, while a low cross-entropy means that the code is easier to predict.

The cross-entropy for the full function names is much higher than in \textit{Hindle et al.}'s work. But it decreases by excluding OOV functions and it gets closer to the cross-entropy they reported on a Java corpus of ten projects. Furthermore, we observe that function names' subtokens have a significantly lower cross-entropy and that excluding the OOV functions has no impact. The no decreasing of the cross-entropy when excluding OOV functions means that almost all subtokens in the test projects appear in the training set. This means that sequences of functions subtokens are more predictable than sequences of full function names. %\textit{We conclude that the naturalness hypothesis is more prevalent using subtokens of function names,} \textbf{but, we may loose important information about the sharing of semantics across functions.}
\textit{We conclude that the naturalness hypothesis is more prevalent using subtokens of function names.} \textbf{However, relaying on only subtokens may make us lose important information about the sharing of semantics across functions. For this reason, we will use the full names to answer RQ2 and RQ3.} 

In their work, \textit{Hindle et al}. estimated $n$-gram models on a Java corpus that includes all tokens present in the code. \textit{Rahman et al}. addressed the same replication work and conclude that syntax tokens are much more present than identifiers in programming languages and that they make the code artificially predictable. The levels of cross-entropy that we report are closer than those reported in \textit{Rahman et al}.'s work. That is, including only functions as training data drastically decreases the predictability of the code.

\begin{figure}[!ht]
    \centering
    \includegraphics[width=\linewidth]{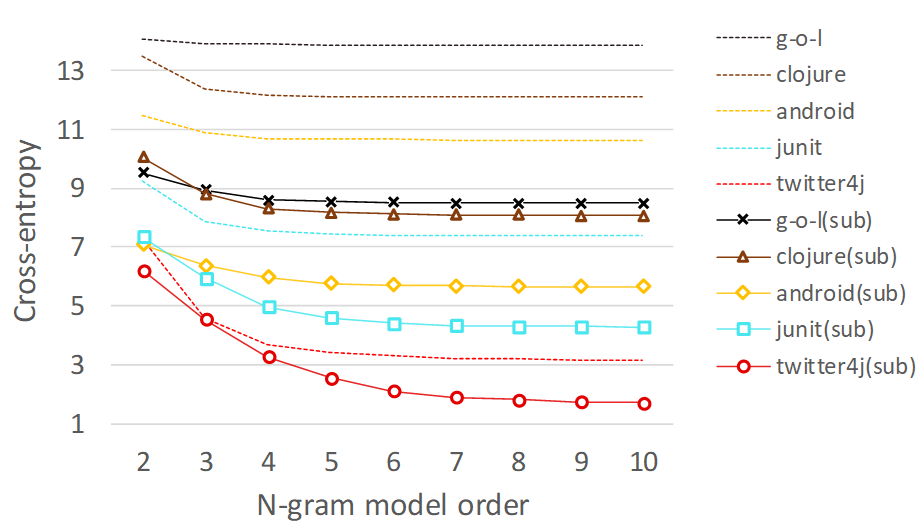}
    \caption{Comparison of the cross-entropy for 5 test projects with full function names and names' subtokens with respect to the order of the $n$-gram model.}
    \label{fig:ngram_10projects}
\end{figure}

To illustrate our words, we report, in Figure \textbf{\ref{fig:ngram_10projects}},  the cross-entropy on 5 test projects using full function names and their subtokens. We observe that the subtokens approach yields a decreasing of the cross-entropy for all test projects. In addition to that, we can observe that some projects such as \textit{twitter4j} and \textit{junit} have a very low cross-entropy, even when considering full names. This can be explained by the high vocabulary coverage of these two projects (\textit{see Table \textbf{\ref{tab:test_set}}}). \textbf{Therefore, we suspect that the paragraph vector and $n$-gram models will perform well on projects that have a high vocabulary coverage.}

\begin{framed}
 To answer RQ1, we have shown that the sequences of function calls included in our training set are difficult to predict by replicating \textit{Rahman et al}'s work \cite{naturalness_revisited}. This also applies to most of our test projects indicating that the task of predicting the next function call with the full name is particularly difficult for these projects.
\end{framed}

% \hs{We should highlight here the answer to RQ1 }
% \mw { I agree, we could use a frame for each RQ as a summary }

%% file: content/5_2_pvmodel.tex
\label{sec:exp_pvmodel}

\begin{figure*}[!ht]
    \centering
    \includegraphics[width=.9\textwidth]{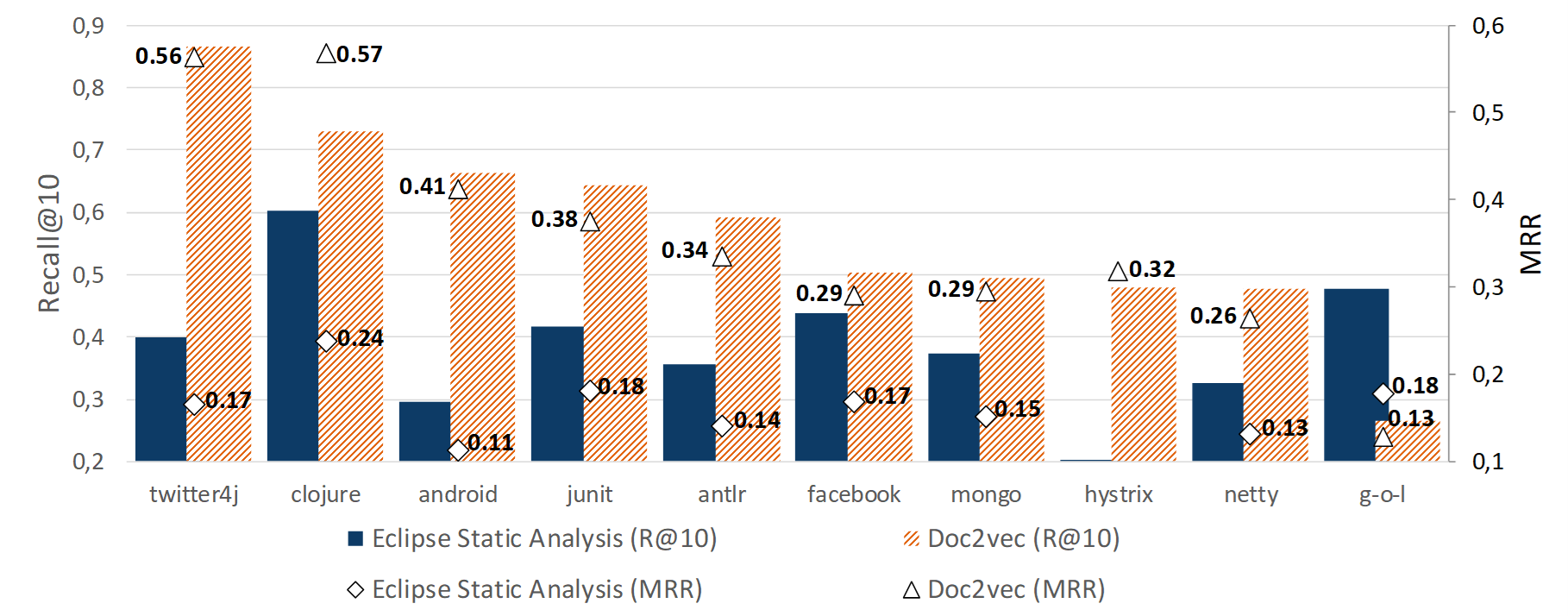}
    \caption{Comparison of Recall@10 and MRR on 10 test projects with Eclipse Static Analysis and Doc2vec.}
    \label{fig:rq2}
\end{figure*}

We compare the performance of the ranking of Eclipse's static analysis candidates with the paragraph vector model and the unranked Eclipse's static analysis. Figure \textbf{\ref{fig:rq2}} summarizes the scores in term of Recall@10 and MRR for both systems on the 10 bold test projects of Table \textbf{\ref{tab:test_set}}. For 9 out of the 10 projects under test, the results highlight a common trend, that is big improvements of Eclipse static analysis when ranking it with the paragraph vector model. The only exception is \textit{game-of-life} which can be explained by the small size of the project and the low percentage of vocabulary coverage (\textit{64\%}).

Two projects (\textit{twitter4j and clojure}), \textbf{each with more than 13.000 completion sites}, stand out from the others with a MRR above 0.5 indicating that the relevant suggestion is on average between the first and the second position in the list. This can be explained by the high vocabulary coverage in these projects (\textit{respectively 99\% and 94\%}). However, in Figure \textbf{\ref{fig:ngram_10projects}}, \textit{clojure} has also a very high cross-entropy meaning that the sequences of functions in the project are difficult to predict. Despite this, our model is able to find useful similar function sequences to perform accurate completions. Therefore, as we suspected in Section \textbf{\ref{sec:exp_ngrams}}, projects with the highest vocabulary coverage have the highest Recall@10 and MRR. 

Finally, another aspect that we evaluated is the time to produce completion suggestions for a call site. This time is on average between 700 ms and 800 ms, which makes our approach usable in a real programming setting. 

\begin{framed}
    To answer RQ2, based on the large number of tested call sites, we can state that using the paragraph vector model to rank potential call candidates, obtained by static analysis, improved dramatically the correctness of the static analysis tool without a negative impact on the response time.
\end{framed}

%% file: content/5_3_comparison.tex
\label{sec:exp_comparison}

\begin{figure*}[!t]
    \centering
    \includegraphics[width=\textwidth]{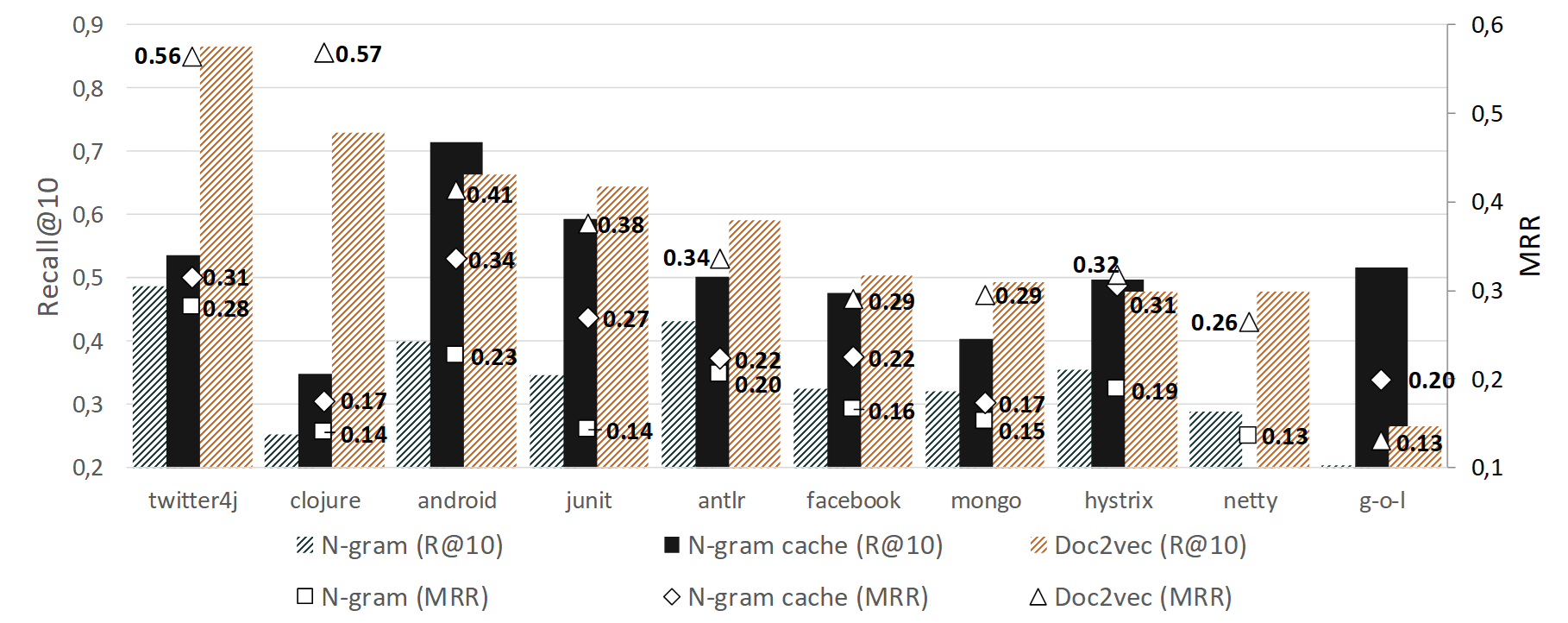}
    \caption{Comparison of Recall@10 and MRR on 10 test projects with $n$-gram, $n$-gram cache and Doc2vec models.}
    \label{fig:rq3}
\end{figure*}

We compare the performance of our model with state-of-the-art $n$-gram language models. Figure \textbf{\ref{fig:rq3}} summarizes the scores and the Table \textbf{\ref{tab:full_results}} shows the overall results for comparison with RQ2. 

As we can observe, the $n$-gram model is the worst performing model. It improves Eclipse's static analysis for only 4 test projects and the performance is far from being of the same order of magnitude as the PV model. Even for projects that have low-level of cross-entropy (\textit{twitter4j and junit, see Figure \textbf{\ref{fig:ngram_10projects}}}), the $n$-gram model is not able to provide accurate recommendations. This is particularly reflected in the low MRRs for these projects, which indicates that the correct recommendation does not appear at the top of the list, on average. 

Given the poor performances of the $n$-gram model, we investigate the use of a $n$-gram model augmented with a cache component. The cache allows to improve considerably the performance of the $n$-gram model. Nevertheless, although the model stores cache information about the project under test during the evaluation, the $n$-gram model is less efficient than the PV model for a great majority of the test projects, as depicted in Figure \textbf{\ref{fig:rq3}}. In addition to that, the $n$-gram model still produces much lower MRRs than the PV model in all the cases but one. For a fairer evaluation, our model should also have included a cache component. Nonetheless, even though we did not implement that mechanism for the PV model, it largely outperforms a $n$-gram cache model that has shown to outperform some LSTM-based deep neural network in \textit{Hellendoorn and Devanbu'}s previous work on source code modeling \cite{ngram_cache}. 

In term of memory footprint and time for completion, the $n$-gram model has a lower computational cost and is faster than the paragraph vector to produce a suggestion. Therefore, despite the promising results, there is room for improvement to produce suggestions faster for the PV model.

\begin{oframed}
To answer RQ3, we can state that the paragraph vector model performs much more effectively than $n$-gram models for the function-call completion task. The PV model is better at abstracting the recurring function sequences in the training set in order to provide insightful recommendations. Therefore, we believe that the usage of models, such as a PV model, that are able to learn more complex relationships between tokens in a sequence (\textit{e.g., non-sequential relationships}) should be favoured for such a task.
\end{oframed}
 
\input{tables/results}

%% file: tables/results.tex
\begin{table*}
\caption{Global results of RQ2 and RQ3 (bold: best, underlined: second best, R@10 in percentage).}
\begin{center}
    \renewcommand{\arraystretch}{1.6}
    \setlength{\arrayrulewidth}{.5pt}
    
    \begin{tabular*}{\textwidth}{@{\extracolsep{\fill}}*{10}{ll|rrrrrrrr}}
    \toprule
    
    \multicolumn{2}{}{} & \multicolumn{2}{c}{\textbf{Eclipse}}  & \multicolumn{2}{c}{\textbf{Doc2vec}} &
    \multicolumn{2}{c}{\textbf{$n$-gram}} &
    \multicolumn{2}{c}{\textbf{$n$-gram cache}} 
    \\ 
    \cline{3-4} \cline{5-6} \cline{7-8} \cline{9-10}
    
    \multicolumn{1}{l}{\textbf{PROJECT}} & \multicolumn{1}{c}{\textbf{Size}} & \multicolumn{1}{c}{R@10} & \multicolumn{1}{c}{MRR} & \multicolumn{1}{c}{R@10} & \multicolumn{1}{c}{MRR} & \multicolumn{1}{c}{R@10} & \multicolumn{1}{c}{MRR} & \multicolumn{1}{c}{R@10} & \multicolumn{1}{c}{MRR}  \\ 
    \midrule
    
        \textbf{game-of-life} & 128 & \underline{47.66} & \underline{0.18} & 26.59 & 0.13 & 20.31 & 0.06 & \textbf{51.56} & \textbf{0.20} \\
        \textbf{android-async-http} & 675 & 29.66 & 0.11 & \underline{66.37} & \textbf{0.41} & 39.85 & 0.23 & \textbf{71.56} & \underline{0.34} \\
        \textbf{facebook-android-sdk} & 5689 & 43.79 & 0.17 & \textbf{50.40} & \textbf{0.29} & 32.41 & 0.16 & \underline{47.65} & \underline{0.22} \\
        \textbf{hystrix} & 5790 & 20.38 & 0.08 & \underline{47.89} & \textbf{0.32} & 35.37 & 0.19 & \textbf{49.67} & \underline{0.31} \\
        \textbf{junit} & 8144 & 41.69 & 0.18 & \textbf{64.44} & \textbf{0.38} & 34.55 & 0.14 & \underline{59.23} & \underline{0.27} \\
        \textbf{antlr} & 11053 & 35.61 & 0.14 & \textbf{59.16} & \textbf{0.34} & 43.04 & 0.21 & \underline{50.17} & \underline{0.22} \\
        \textbf{clojure} & 13020 & \underline{60.28} & \underline{0.24} & \textbf{72.90} & \textbf{0.57} & 25.16 & 0.14 & 34.88 & 0.17 \\
        \textbf{twitter4j} & 13365 & 39.96 & 0.17 & \textbf{86.65} & \textbf{0.56} & 48.71 & 0.28 & \underline{53.47} & \underline{0.31} \\
        \textbf{mongo-java-driver} & 35836 & 37.34 & 0.15 & \textbf{49.37} & \textbf{0.30} & 32.00 & 0.15 & \underline{40.31} & \underline{0.17} \\
        \textbf{netty} & 72754 & 32.58 & 0.13 & \textbf{47.82} & \textbf{0.26} & 28.95 & 0.13 & - & - \\
        
    \bottomrule
    \end{tabular*}
\end{center}

\label{tab:full_results}
\end{table*}

%% file: content/5_4_threats.tex
\label{sec:threats}

We identified some threats to the validity of our evaluation and attempted to address them during its design. 
The first threat relates to the mono-operation bias as we experimented only with Java projects. We conjecture that our approach can be used for call completion in other languages as we do not rely on Java language constructs, but on identifiers. 
To prevent the mono-method bias, we evaluated our approach with two metrics commonly used to measure the effectiveness of ranking systems, R@k and MRR. 
Another threat concerns the interaction of setting and treatment. Indeed, we reused and compared our results with the completion in Eclipse. It has been shown that Eclipse's static analysis tools are commonly used by Java developers \cite{Murphy_eclipse}, and we do believe that it is representative enough. Similarly, we compared our paragraph vector model with only one other type of model, i.e., $n$-gram model. Nevertheless, we considered two variants of this model and believe that it provides a good basis for comparison between two kind of models that are completely different in nature and widely used in the literature.
Another important aspect that we considered is the representativeness of the dataset. We made sure to train our models on a large dataset of open-source projects from different domains of application and of variable sizes. For the evaluation, we choose a variety of test projects as well. 
Finally, an important threat to the validity of our results arises from the choice of hyper-parameters of the paragraph vector models. To address the issue, we followed guidelines from the literature. We tuned the hyper-parameters that influence the most the quality of embeddings\footnote{https://code.google.com/archive/p/word2vec} and chose commonly used values for the other hyper-parameters, following Lau and Baldwin's recommendations \cite{doc2vec_empirical}. We reported the chosen values in Section \textbf{\ref{sec:eval_pv}}.

%% file: content/6_related_work.tex
\label{sec:related_work}

Neural approaches, $n$-gram and embedding-based language models have been widely used for automating tasks of the software development lifecycle. However, we focus on code completion by contrasting previous works with our approach. Then, we discuss about source code modeling and the broader usage of embedding-based approaches on source code (\textit{see the literature study by Chen and Monperrus \cite{study_embeddings} and the survey by Allamnis et al. \cite{allamanis_survey} for more references on these topics}).

\subsubsection*{\textbf{Code Completion}}
Code completion has been an active field of research in software engineering. In one of the ealier learning-based approaches, Bruch et al. \cite{bruch2009} used k-nearest-neighbors to find relevant code suggestions using features extracted from the call site. Later, Proksch et al. \cite{proksch2015} improved their work by using Bayesian networks and gathering more context information. The main limitation of these techniques is that they are designed to predict calls of particular APIs and require training data specific to these APIs. In our embedding-based approach, the model learns distributed representations of the source code from a large training corpus and does not require to extract manual information from the project under development. With the hypothesis of \textit{naturalness of software}, Hindle et al. \cite{naturalness} outlined the possibility to use $n$-gram language models for code completion by predicting a call given the previous code tokens. Tu et al. \cite{localness} used cache $n$-gram language models for code completion by capturing local patterns in the code. Hellendoorn and Devanbu \cite{ngram_cache} extended this approach by improving the cache component with information about the scope of the call site. Nguyen et al. \cite{nguyen_2013} proposed an extension of $n$-gram language models by incorporating semantic information about the completion context. Similarly, Nguyen et al. \cite{gralan} used AST-based language models to learn higher-level patterns than $n$-gram language models to improve API code suggestion. Raychev et al. \cite{raychev2014} compared the performance of $n$-gram and neural language models for Android API code suggestion. In this work, we compared our approach with two configurations of $n$-gram models and show that a paragraph vector model is able to outperform both $n$-gram models on function-call completion. Moreover, the chosen $n$-gram baselines are strong since they have shown to be efficient and sometimes a better choice than RNN/LSTM-based neural networks for source code modeling \cite{ngram_cache}. We also show that such models can be integrated with a type-based completion tool to suggest only tangible function calls. 

Recent approaches using deep learning have mainly focused on learning representations of ASTs with attention-based neural networks. Bhoopchand et al. \cite{bhoopchand2016} used pointer networks to learn long-range dependencies in Python ASTs for identifiers completion. Li et al. \cite{li2017} used the same approach with a focus on out-of-vocabulary identifiers. Liu et al. \cite{Liu2017NeuralCC} leveraged LSTM neural networks trained on partial ASTs to predict nodes in a target AST. Karampatsis et al. \cite{karampatsis2020big} proposed a LSTM neural networks that is able to suggest out-of-vocabulary identifiers by learning the internal structure of code tokens. 
In the same vein, Svyatkovskiy et al. \cite{Svyatkovskiy_2019} compared several neural network architectures for method and API recommendations in Python. They learn AST-based representations of code snippets to perform the completion by comparing a call site context with the representations learned by their model. In a subsequent paper, Svyatkovskiy et al. \cite{svyatkovskiy2020fast} defined a framework using the same approach combined with an existing code completion tool to produce ranked lists of suggestions. Alon et al. \cite{alon2019structural} proposed an approach where a transformer model learns to predict an AST node given all possible AST paths leading to this node. Finally, \textit{Kim et al.} \cite{kim2020code} designed the same kind of approach but compared several ways to feed AST trees into a transformer model and focused the evaluation of their model on predicting specific types of tokens. These deep learning based works have shown to be efficient for code completion and especially to predict common API calls. In future works, we plan to compare our approach with deep neural network approaches for the completion of function-calls in general and not specifically for particular APIs.

\subsubsection*{\textbf{Source Code Modeling}}
Recent works on source code modeling have focused on learning probabilistic models of code. Source code modeling is usually an upstream task of predictive tasks such as code completion.
Approaches based on $n$-gram language models have shown to be useful to find regularities in code \cite{naturalness, nguyen_2013, localness, ngram_cache}. More recent approaches are based on distributed representations of source code \cite{dist, word2vec_1, word2vec_2, doc2vec} that learn more complex semantic relationships between code tokens. Both kinds of approaches can be useful for some downstream tasks.
Allamanis et al. \cite{Allamanis_suggestion} used embedding-based language model to predict method names. Nguyen et al. \cite{api_mapping} learned embeddings of API elements and try to map them across programming languages. Gu et al. \cite{api_code_search} proposed an embedding-based approach to find relevant API sequences given a search query. From another paerspective, White et al. \cite{prog_repair_1} and Chen and Monperrus \cite{prog_repair_2} used embeddings to find similarities in code for automatic program repair. Finally, Büch and Andrzejak \cite{clone_detection} learned embedding of ASTs of methods for clone detection. These previous works show a broad range of applications in which our embedding-based approach could be used with small adaptations. These include method names prediction \cite{Allamanis_suggestion, code2seq}, clone detection \cite{embedding_clone_detection, embedding_clone_detection_2}, API pattern detection \cite{embedding_api} or code search \cite{aroma2018, focus, dl_code_search, husain2020codesearchnet}.

%% file: content/7_conclusion.tex
\label{sec:conclusion}

% {\color{blue}
% 
% }

In this paper, we presented an approach for function-call completion that can be integrated with a static analysis tool based on a language typing system. Our approach starts from the assumption that it is possible to abstract application-independent high-level concepts in the form patterns of call sequences contained in code repositories. To this end, we build on document-embedding algorithms to train models that can be exploited for function-call completion.
Our experiments highlights promising results for most of the tested projects and indicate that our trained model captures useful high-level concepts that can be used for completion. This shows that our approach can be useful for helping developers writing their software even for new projects and with limited knowledge about the used APIs.

Although the obtained results are satisfactory, there is room for improvement. One of the limitations of our approach is that it is less efficient with projects having very specific function names, not frequent in existing code repositories. We plan to improve the natural-language processing pipeline to cope with this situation. We also plan to explore other embedding-based language models to improve the completion. Finally, instead of capturing high-level concepts inside a method scope, we plan to learn similar concepts in wider scopes and thus learning recurring long-range dependencies that could be useful for program summarizing, for instance.

From another perspective, the fact that our approach does not rely on language constructs, but rather on sequences of identifiers used in method names opens the door for many other possibilities to explore. Indeed, we conjecture that the learned models can be reused cross-programming languages. They can also be used, with some adaptation, to assist developers for other tasks such as program documentation by providing summaries, construct naming for automated generation, clone detection, and code search. Finally, an approach similar to ours can be employed to assist in building design diagrams such those of UML.